\documentclass[aps,prl,twocolumn,superscriptaddress,notitlepage,groupedaddress]{revtex4-1}

\usepackage{graphicx}
\usepackage{mathptmx, textcomp}
\usepackage[usenames]{xcolor}

\usepackage{amsmath}
\usepackage[colorlinks,urlcolor=blue,citecolor=blue,linkcolor=blue]{hyperref}
\usepackage{cleveref}
\Crefname{figure}{Fig.}{Figs.}
\Crefname{equation}{Eq.}{Eqs.}
\Crefname{table}{Tbl.}{Tbls.}
\usepackage{siunitx}
\usepackage{physics}
\usepackage{hyperref}
\usepackage{tabularx}
\newcolumntype{Y}{>{\centering\arraybackslash}X}
\usepackage[notextcomp]{stix}
\usepackage{oplotsymbl}

\newcommand{\1}{|1\rangle}
\newcommand{\2}{|2\rangle}
\newcommand{\K}{$^{39}$K}
\newcommand{\br}{\textbf{r}}

\begin{document}

\title{Observation of a Lee-Huang-Yang Fluid}

\author{Thomas G. Skov}
\author{Magnus G. Skou}
\author{Nils B. J{\o}rgensen}
\author{Jan J. Arlt}

\affiliation{Center for Complex Quantum Systems, Department of Physics and Astronomy, Aarhus University, Ny Munkegade 120, DK-8000 Aarhus C, Denmark.}
%\date{\today}

\begin{abstract}
We observe monopole oscillations in a mixture of Bose-Einstein condensates, where the usually dominant mean-field interactions are canceled. In this case, the system is governed by the next-order Lee-Huang-Yang (LHY) correction to the ground state energy, which describes the effect of quantum fluctuations. Experimentally such a LHY fluid is realized by controlling the atom numbers and interaction strengths in a \K~spin mixture confined in a spherical trap potential. We measure the monopole oscillation frequency as a function of the LHY interaction strength as proposed recently by Jørgensen \emph{et al.} [\href{https://doi.org/10.1103/PhysRevLett.121.173403}{Phys. Rev. Lett.
	121, 173403 (2018)}] and find excellent agreement with simulations of the complete experiment including the excitation procedure and inelastic losses. This confirms that the system and its collective behavior are initially dominated by LHY interactions. Moreover, the monopole oscillation frequency is found to be stable against variations of the involved scattering lengths in a broad region around the ideal values, confirming the stabilizing effect of the LHY interaction. These results pave the way for using the non-linearity provided by the LHY term in quantum simulation experiments and for investigations beyond the LHY regime.
\end{abstract}

\maketitle

The advent of well controlled mixtures of quantum gases with tunable interaction strengths has enabled fascinating insights beyond the mean-field description of such systems. This is particularly important for strongly interacting systems of current interest, where the mean-field description typically breaks down as higher-order effects become important. Most prominently, the next-order Lee-Huang-Yang (LHY) correction describes the effect of quantum fluctuations on the ground-state energy of a bosonic quantum gas \cite{Lee1957}, and its effect has been observed in several experiments \cite{Altmeyer2007,Shin2008,Papp2008,Navon2010,Navon2011} emphasizing its importance in describing systems beyond the mean-field regime.

The ability to tune the interaction strength is especially useful in cases with competing interactions, such as two-component quantum mixtures where both inter- and intracomponent interactions are relevant. Here, the LHY contribution to the energy functional has been extended to the case of homonuclear  bosonic mixtures \cite{Larsen1963} and recently an effective expression was derived for the heteronuclear case \cite{Minardi2019}. In particular, the influence of LHY physics can be observed more readily by tuning the interactions such that other contributions to the energy density are suppressed. This approach enables the formation of self-bound droplets stabilized by the repulsive LHY energy contribution \cite{Petrov2015}, which have been observed and investigated in homonuclear \cite{Cabrera2018,Semeghini2018,Cheiney2018,Ferioli2019} and heteronuclear \cite{DErrico2019} bosonic mixtures. Similar observations have been made in dipolar quantum gases \cite{Kadau2016,FerrierBarbut2016,Schmitt2016,Chomaz2016} culminating in the observation of supersolid behavior in these systems \cite{Boettcher2019,Chomaz2019,Tanzi2019}.

Here, we consider a quantum mixture where the mean-field interactions cancel exactly such that interactions in the mixture are governed primarily by the LHY correction \cite{Jorgensen2018}. In practice, this can be achieved utilizing a two-component Bose-Einstein condensate (BEC) characterized by scattering lengths $a_{ij}$ between components $i$ and $j$. For scattering lengths $\delta a = a_{12} + \sqrt{a_{11}a_{22}} = 0$ and densities $n_2/n_1 = \sqrt{a_{11}/a_{22}}$, the mean-field contributions to the energy functional vanish, and the resulting LHY fluid can be characterized by measuring its monopole oscillation frequency, which differs significantly from that of a single-component BEC \cite{Jorgensen2018}.

\begin{figure}[b!]
	\centering
	\includegraphics[width=\columnwidth]{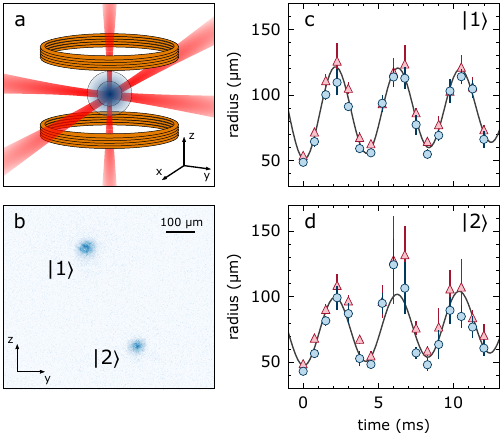}
	\caption{(a) Schematic representation of the experiment showing a LHY fluid undergoing monopole oscillations in a spherical potential composed of three red-detuned laser beams. (b) Typical absorption image after time-of-flight (TOF) during which atoms in states $\1$ and $\2$ are separated by a magnetic field gradient. (c-d) Extracted BEC radii of atoms in states $\1$ (c) and $\2$ (d) after 28 ms TOF as a function of evolution time. The radii along the $y$- and $z$-directions are shown as blue dots and red triangles, respectively, and the grey lines are fits of \Cref{eq:fitmodel} to the mean radius. The data was recorded for $\omega_0 = 2 \pi \times \SI{113.1(6)}{Hz}$ and $N = \num{2.3(3)e4}$ corresponding to $U = 0.5(1)$.} 
	\label{fig:exp}
\end{figure}

In this letter we experimentally investigate the collective excitations of such a LHY fluid in  a $^{39}$K spin mixture. We measure the monopole oscillation frequency depending on the strength of the LHY interaction and its magnetic field dependence in the vicinity of the $\delta a = 0$. To evaluate the results, we perform detailed simulations of the system including the effect of inelastic losses and the preparation sequence of the mixture. Throughout the investigated region we find good agreement between experimental observations and theory demonstrating a detailed understanding of the LHY fluid.

To realize a LHY fluid experimentally, we employ the $|F = 1, m_F = -1 \rangle \equiv \1$ and $|F = 1, m_F = 0 \rangle \equiv \2$ states of the lowest hyperfine manifold of $^{39}$K, which offer favorable Feshbach resonances for realizing $\delta a = 0$ \cite{Petrov2015, Cabrera2018,Semeghini2018, Jorgensen2018}. Based on models for the scattering lengths presented in Refs. \cite{Tanzi2018,Chapurin2019,Cheiney2018} we find that the system
fulfills $\delta a = 0.0(3)$ at 56.83(2) G with $a_{11} = 33.3(3)\: a_0$, $a_{22} = 84.2(3)\:a_0 $, and $a_{12} = -52.97(1)\:a_0$,  where $a_0$ is the Bohr radius. Given these scattering lengths, the requirement for the relative densities is $n_2/n_1 = 0.629(3)$ corresponding to $\sim 40\:\%$ of the total atom number in the $\2$ state.

In practice, the experiment starts from a nearly pure BEC in the $\1$ state \cite{Wacker2015} trapped in a spherically symmetric harmonic potential. The employed optical dipole trap (ODT) is composed of a beam along each Cartesian axis (\Cref{fig:exp}(a)) allowing the realization of almost identical trap frequencies, $\omega_0$, in all directions. The initial BEC in the $\1$ state is prepared at the target magnetic field in the vicinity of  $\delta a = 0$. Subsequently, the measurement is initialized by transferring part of the atoms to the $\2$ state using a radio-frequency (rf) pulse tuned to the bare atomic transition. Due to the sudden change in the interaction strengths, the system starts to contract and strong monopole oscillations are initialized. This preparation method is necessary since inelastic losses of atoms in the $\2$ state limit the lifetime of the mixture \cite{Cabrera2018,Semeghini2018,Cheiney2018,Ferioli2019}. Afterwards, the mixture is held for a variable evolution time before release from the trap and subsequent absorption imaging. During time-of-flight (TOF), the states are separated by applying a magnetic field gradient, which enables the resulting cloud profiles to be evaluated separately  (\Cref{fig:exp}(b)). Note that imaging the clouds after TOF results in a phase shift of $\pi$ compared to the in-trap oscillations since the momentum distribution is observed.

For each evolution time, the cloud profiles are fitted using a Thomas-Fermi profile and the cloud radii along the $y$- and $z$-directions are extracted. An example of such a measurement is shown in \Cref{fig:exp}(c-d), where the cloud radii feature large amplitude oscillations as a consequence of the experimental preparation method. For both states, the $y$- and $z$-radii oscillate in phase confirming that the oscillations are monopolar. Moreover, the two clouds initially oscillate jointly as expected for a LHY fluid, which can be described in a one-component framework \cite{Jorgensen2018}. With increasing evolution time, however, inelastic losses of atoms in the $\2$ state result in the appearance of small phase differences and deviations from pure sinusoidal behavior. Due to these losses, we use the radius of the cloud in the $\1$ state for the further evaluation since it provides more reliable results.

The large oscillation amplitude and inelastic losses require a detailed theoretical analysis, and we simulate the experiment \cite{Antoine2014,Antoine2015} using two coupled Gross-Pitaevskii equations including the LHY contributions \cite{Larsen1963, Petrov2015}. We first calculate the ground state of a pure BEC in the $\1$  state at the target magnetic field. The fast transfer is then modeled by assuming that both components start in the calculated ground state wave function, but with properly adjusted atom numbers and scattering lengths. In agreement with experiment, this results in monopole oscillations of large amplitude. Inelastic losses due to three-body recombination are included using imaginary terms corresponding to the relevant loss channels \cite{SM}.

\begin{figure}[t!]
	\centering
	\includegraphics[width=\columnwidth]{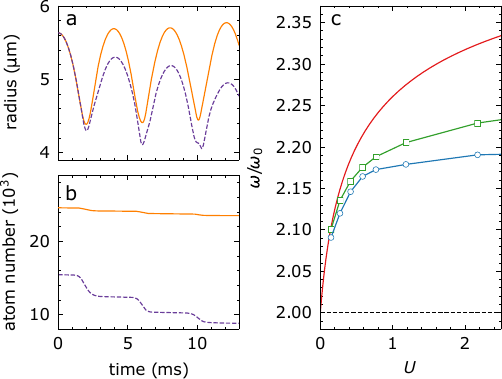}
	\caption{Results of the dynamical two-component simulations. (a-b) Simulated radii and atom numbers for a LHY fluid including inelastic losses for $\omega_0 = 2 \pi \times \SI{113.6}{Hz}$ and $N = \num{4e4}$ corresponding to $U = 1.2$. Results for atoms in states $\1$ and $\2$ are shown as solid orange and dashed purple lines, respectively. (c) Extracted oscillation frequencies of atoms in the $\1$ state as a function of LHY interaction strength. Results with and without inelastic losses are shown as blue circles and green squares, respectively. The low-amplitude limit is shown as a red line and the non-interacting limit is shown as a dashed black line.}
	\label{fig:sim}
\end{figure}

This simulation allows us to quantify the monopole oscillation frequency as a function of the dimensionless parameter $U = N^{3/2} |a_{12}/a_{\rm{ho}}|^{5/2}$ with total atom number $N$, harmonic oscillator length $a_{\rm{ho}} = \sqrt{\hbar/m\omega_0}$ and mass $m$. This parameter corresponds to the ratio of the LHY interaction energy to the kinetic energy and thus characterizes the interaction strength of the LHY fluid \cite{Jorgensen2018}. Figure \ref{fig:sim}(a) shows simulated cloud radii for $U = 1.2$ including inelastic losses based on the three-body loss coefficients given in Refs. \cite{Cheiney2018,Semeghini2018}. Similar to the experiment, the two components initially contract and start to oscillate jointly. As the density increases, the losses in the $\2$ state set in, resulting in a cascading loss of atoms coinciding with the minima of the radii (\Cref{fig:sim}(b)), leading to a decay of the oscillations. Since the losses primarily affect atoms in the $\2$ state, the system is displaced away from the ideal density ratio $n_2/n_1 = \sqrt{a_{11}/a_{22}}$ resulting in increasing mean-field interactions. This leads to an additional repulsion of atoms in the $\1$ state, which explains the increasing offset and amplitude visible in their cloud radius. As a consequence, the resulting oscillation is determined both by the initial evolution, governed by the dominant LHY correction, and the later evolution, where the mean-field contributions become relevant and the system deviates from a pure LHY fluid.

To capture these effects, we fit a model function to the simulated radius, $r$, as a function of evolution time, $t$,
\begin{align}
	r(t) = r_0 + st + A\sin(\omega t + \phi) \exp(-t/\tau).
	\label{eq:fitmodel}
\end{align}
Here, $r_0$ is an offset radius, $s$ is the slope, $A$ is the oscillation amplitude, $\omega$ is the angular frequency, $\phi$ is a phase offset, and $\tau$ is the time constant describing the growth or decay of the oscillations.
Note that a small frequency chirp could arise during the evolution time due to the increasing mean field interactions. In our simulations, however, we do not observe a significant chirp and therefore neglect it. As a result, the extracted frequency $\omega$ describes the average oscillation frequency within the data range. Even though the simulated radius deviates from a regular sinusoidal in its extrema as shown in \Cref{fig:sim}(a), \Cref{eq:fitmodel} captures the oscillation frequency well.

Figure \ref{fig:sim}(c) shows the simulated monopole oscillation frequency of atoms in the $\1$ state as a function of the LHY interaction strength, with and without inelastic losses. For comparison, the oscillation frequency of the LHY fluid in the low-amplitude limit \cite{Jorgensen2018} and the non-interacting limit are shown. For low interaction strengths, all results including the LHY correction follow a common rising trend, however our simulations quickly deviate from the low-amplitude result showing a pronounced reduction in frequency as a consequence of the large oscillation amplitudes. Furthermore, the effect of inelastic losses becomes apparent as an additional decrease of the oscillation frequency, settling at $\omega/\omega_0 \sim 2.18$ for the investigated trap \cite{trapnote}. Comparing the simulated results to the non-interacting limit, it is clear that the LHY correction has considerable influence on the oscillatory behavior, even under the influence of inelastic losses. Based on this thorough theoretical analysis, a comparison with our experimental results is now possible.

%-----------------------------------------------------------

In a first set of experiments, we investigate the dependence of the monopole frequency on the LHY interaction strength. We follow the experimental procedure outlined above, preparing the initial BEC at the magnetic field corresponding to $\delta a = 0$ and initialize the mixture using a rf pulse of \SI{1.3}{\micro\second} duration, which realizes the required density  ratio. 
This pulse length was chosen based on Rabi oscillation measurements using cold thermal clouds. To scan the LHY interaction strength, we adjust the total number of initial BEC atoms by 
varying the loading time of $^{39}$K in the dual-species magneto-optical trap \cite{Wacker2015}. The resulting oscillations yield measurements similar to \Cref{fig:exp}(c-d) and we extract the oscillation frequency by fitting \Cref{eq:fitmodel} to the mean of the cloud radii in $y$- and $z$-direction as a function of evolution time \cite{SM}.

\begin{figure}[htbp]
	\centering
	\includegraphics[width=\columnwidth]{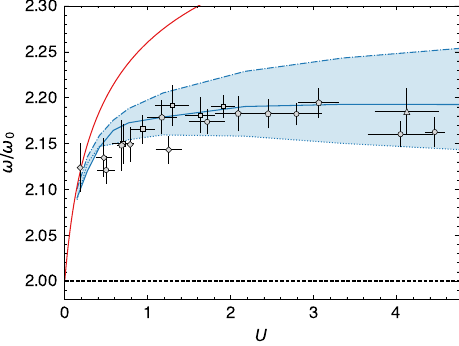}
	\caption{Observed monopole oscillation frequency depending on interaction strength of the LHY fluid for spherical traps with $\omega_0 = 2 \pi \times \SI{113.1(6)}{\Hz}$ (circles), \SI{113.7(1)}{Hz} (squares),  \SI{115.3(4)}{Hz} (diamonds), \SI{111.1(3)}{Hz} (pentagons), and \SI{110.9(5)}{Hz} (triangles). Simulated results calculated for $\omega_0 = 2 \pi \times \SI{113.6}{\Hz}$ including inelastic losses are shown as a blue line. The dash-dotted and dotted lines correspond to simulations without losses and a doubled loss coefficient for the channel involving three atoms in the $\2$ state, respectively. The red line shows the monopole frequency of an ideal LHY fluid in the low-amplitude limit and the dashed black line shows the non-interacting limit.}
	\label{fig:main}
\end{figure}

Figure \ref{fig:main} shows experimental results for a range of trap frequencies together with the simulated results for $\omega_0 = 2 \pi \times \SI{113.6}{Hz}$ also shown in \Cref{fig:sim}. Since the exact loss coefficients are not well known, we include two limiting cases in our simulations: The upper limit of the light blue area shows simulated results neglecting losses entirely and the lower limit doubles the loss coefficient for the channel involving three atoms in the $\2$ state, corresponding to the upper limit given in Refs. \cite{Cheiney2018,Semeghini2018}. For comparison, we again show the low-amplitude limit of a LHY fluid and the non-interacting limit.
The experimentally obtained oscillation frequencies display a clear upward trend for $U \lesssim 1.5$ and for increasing interaction strengths, the oscillation frequency settles at a value determined by the LHY interactions, the large oscillation amplitudes, and inelastic losses. The overall agreement between theory and experiment is very good and we conclude that the mixture is indeed initially dominated by the LHY correction.

In a second set of experiments, we investigate the stability of the monopole frequency of the system against variations of the scattering lengths in the vicinity of $\delta a = 0$. The experiment is performed using a trap frequency $\omega_0 = 2 \pi \times \SI{110.9(5)}{\Hz}$ and atom number $N = \num{9.4(6)e4}$ corresponding to $U = 4.1(4)$ for $\delta a = 0$. We vary the magnetic field at which the mixture is prepared, effectively varying $\delta a$, which is primarily influenced by $a_{22}$, since $a_{11}$ and $a_{12}$ are approximately constant within the investigated magnetic field range. Note that we keep the length of the rf pulse, which initializes the experiment, constant and hence the ideal density ratio, $n_2/n_1 = \sqrt{a_{11}/a_{22}}$ is only fulfilled initially at exactly $\delta a = 0$. Within the range of magnetic fields investigated, this corresponds to a minor relative deviation from the ideal ratio by $\pm 10 \%$, which is included in the simulations.

\begin{figure}[h!]
	\centering
	\includegraphics[width=\columnwidth]{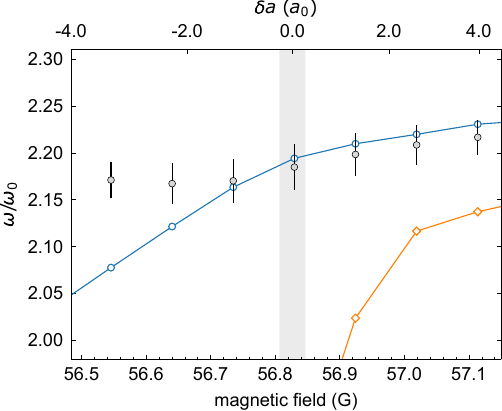}
	\caption{Monopole frequency of the spin mixture depending on magnetic field and $\delta a$. Experimental results for $\omega_0 = 2 \pi \times \SI{110.9(5)}{Hz}$ and $N = \num{9.4(6)e4}$ corresponding to $U = 4.1(4)$ are shown as gray points. Simulated results using the same parameters with and without including the LHY correction are shown as blue circles and orange diamonds, respectively. The shaded region indicates the magnetic field regime where $\delta a = 0$ within $1 \sigma$. The horizontal errors on the experimental results are smaller than the markers.}
	\label{fig:Bfield}
\end{figure}

Figure \ref{fig:Bfield} shows experimentally observed and simulated monopole oscillation frequencies as a function of magnetic field and $\delta a$. When including the LHY correction, the simulated oscillation frequency decreases slowly with decreasing magnetic field. On the contrary, omitting the LHY correction results in a rapid decrease of the oscillation frequency towards the non-interacting limit of $\omega/\omega_0 = 2$ when approaching $\delta a = 0$. Beyond this point, the system collapses without the repulsive energy contribution from the LHY correction.

For $\delta a \gtrsim  0$ the  experimental results agree very well with the simulation including the LHY correction. For negative $\delta a$ the agreement is less good, which we attribute to the sensitivity of the simulation to the exact loss coefficients for increasingly attractive mean-field interactions. Here, the losses almost completely remove the population in state $\2$, which reduces the validity of our simulation. Nonetheless, the experimental data shows that the LHY correction vastly influences the monopole oscillation frequency in a region around $\delta a = 0$ and prevents the collapse of the mixture for attractive mean-field interactions. This agrees with the theoretical results of Ref. \cite{Jorgensen2018} which found that the LHY energy dominates the interactions within a window around the ideal magnetic field. Note that droplet formation in the regime $\delta a <0$ is contained in our theoretical analysis, however our experimental atom number ratios are unfavorable for the observation of droplets.

In conclusion, we have experimentally realized a LHY fluid and investigated its monopole oscillation frequency depending on the LHY interaction strengh, finding excellent agreement with detailed simulations taking the experimental preparation method and inelastic losses into account.
Our experimental results show that the monopole frequency of the system is stable against variations of the involved scattering lengths, displaying a large stability region around $\delta a =  0$, where the repulsive LHY interaction prevents collapse of the mixture.

These results pave the way for further investigations of the LHY dominated regime with interesting research directions including higher-order collective modes and different trap geometries. Furthermore, the LHY fluid provides a promising platform for observing even higher-order effects such as the next-order correction to the energy of a Bose gas, $E_{\rm{WHPS}}$ \cite{Fetter1971, Christensen2015}. On a broader scale, the realization of a LHY fluid enables new quantum simulation experiments utilizing the quartic non-linearity, which governs the interactions of the system. Such experiments would ideally be realized in a system suffering less severely from inelastic losses than the \K ~spin mixture considered here. Building on the results of \cite{Minardi2019}, a promising candidate for such experiments could therefore be the $^{41}$K-$^{87}$Rb mixture, which was recently found to support the existence of long-lived quantum droplets \cite{DErrico2019}.

We acknowledge support from the Villum Foundation, the Carlsberg Foundation, the Independent Research Fund Denmark, and the Danish National Research Foundation through the Center of Excellence ``CCQ'' (Grant agreement no.: DNRF156).

\bibliography{LHYbib_combined}

\begin{thebibliography}{37}
\bibitem [{tra()}]{trapnote}%
  \BibitemOpen
  \href@noop {} {}\bibinfo {note} {We have confirmed that varying the trap
  frequency in the range 111-118 Hz makes negligible difference in the
  extracted frequency.}\BibitemShut {Stop}%
\end{thebibliography}

%-----------------------------------------------------------
%\newpage

\section{Supplemental Material}

\subsection{Trap frequency determination}
To realize an optical dipole trap (ODT) with equal trap frequencies along all Cartesian axes, an additional laser beam with a wavelength of 1064 nm and a waist of \SI{61}{\micro\meter} was added along the vertical ($z$) axis in addition to the ODT beams described in Ref. \cite{Wacker2015}. 
To measure the resulting trap frequencies, we prepare a BEC in the $\1$ state in the desired trap configuration. The powers in all beams are abruptly increased by  $\sim \SI{10}{\percent}$ for 1 ms, after which the powers are once again decreased to the values giving the desired trap potential. Following a variable evolution time, the BEC is subsequently released from the trap and imaged after 28 ms time-of-flight (TOF). To measure the trap frequencies along all Cartesian axes, separate measurements are carried out with absorption imaging along the $x$- and $z$-directions respectively. Due to imperfections in the optical setup, the axes of the trapping potential lie in a frame $(x', y', z')$, which is rotated with respect to the reference frame ($x, y, z$) set by the imaging axes. The resulting oscillation data measured in the ($x, y, z$) frame thus corresponds to a beat signal between the eigenmodes of frequency $\omega_{i}'$, $i = x, y, z$, in the reference frame of the trap, which can be obtained by implementing a rotation of the coordinate system in the analysis \cite{Altmeyer2007,Scherer2013,Lobser2015}. In practice, we assume that the center-of-mass motion in the reference frame of the trap can be described by
\begin{align}
	\mathbf{r}' = \begin{bmatrix}
		A_x \sin(\omega_x't+\phi_x) \\
		A_y \sin(\omega_y't+\phi_y)\\
		A_z \sin(\omega_z't+\phi_z)
	\end{bmatrix},
\end{align}
where $A_i$, $\omega_i'$, and $\phi_i$ are the amplitudes, angular frequencies, and phase offsets of the oscillations. In the reference frame defined by the imaging setup, the position of the atomic cloud is then given by
\begin{align}
	\mathbf{r} = \mathbf{R}_z(\theta_z) \mathbf{R}_y(\theta_y) \mathbf{R}_x(\theta_x) \mathbf{r}',
	\label{eq:rotated_coords}
\end{align}
where $\mathbf{R}_i(\theta_i)$ is the basic rotation matrix which rotates the coordinate system around axis $i'$.

\begin{figure*}[]
	\centering
	\includegraphics[width=\linewidth]{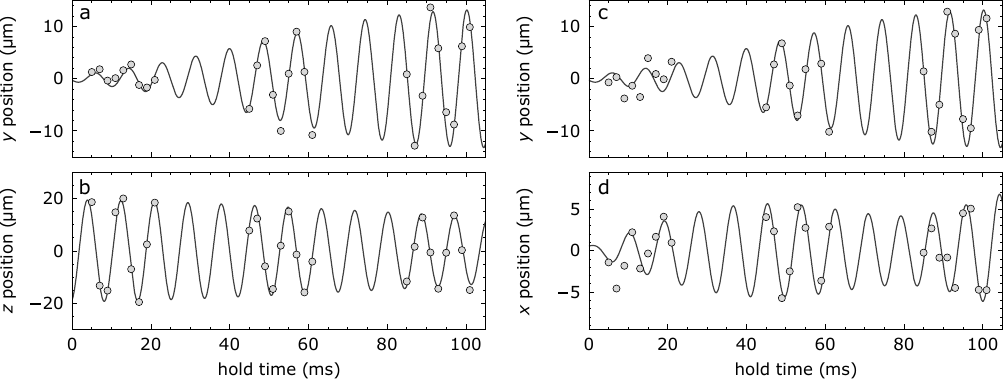}
	\caption{Trap frequency measurement corresponding to the square markers in Fig. 3 of the main text. Positions obtained from imaging along the $x$- (a-b) and $z$-axes (c-d) are shown together with a simultaneous fit of \Cref{eq:rotated_coords} to both imaging directions. Due to the destructive imaging technique, the $y$-positions are different within the two data sets.} 
	\label{fig:trap_freqs}
\end{figure*}

To extract the trap frequencies $\omega_{i}'$, we simultaneously fit \Cref{eq:rotated_coords} to the cloud positions measured using the  $x$- and $z$-imaging systems. Figure \ref{fig:trap_freqs} shows example data for the trap corresponding to the square markers in Fig. 3 of the main text. Note, that due to the destructive imaging technique we only image along one direction per experimental sequence. The fit of \Cref{eq:rotated_coords} to the data yields trap frequencies $\omega_x' = 2 \pi \times \SI{118.5(1)}{\Hz}$, $\omega_y' = 2 \pi \times \SI{115.3(2)}{\Hz}$, and $\omega_z' = 2 \pi \times \SI{108(1)}{\Hz}$, and rotation angles $\theta_x = \SI{101(4)}{\degree}$, $\theta_y = \SI{120(2)}{\degree}$, and $\theta_z = \SI{-125(3)}{\degree}$. Such large rotation angles are not unexpected due to the small difference between the trap frequencies.

The three frequencies are combined into a geometric mean,  $\omega_0 = (\omega_x'\omega_y' \omega_z')^{1/3}$, which is used in the analysis. In \Cref{tbl:trap_freqs} we show the fitted values of $\omega_0$ together with the frequency differences with respect to the individual axes. Note that achieving a perfectly spherical trap is an experimental challenge involving an iterative adjustment of the powers in the trapping beams and measurements of the trap frequencies.

\begin{table*}
	\begin{tabularx}{\linewidth}{l c Y Y Y Y}
		\hline
		Data set &  & $\omega_0/2\pi$ (Hz) & $(\omega_x'-\omega_0)/2\pi$ (Hz) & $(\omega_y'-\omega_0)/2\pi$ (Hz) & $(\omega_z'-\omega_0)/2\pi$ (Hz) \\
		\hline\hline
		Fig. 3 & $\circlet$ & \num{113.1(6)} & \num{3(1)} & \num{0.1(9)} & \num{-3.3(8)} \\
		& $\mdwhtsquare$ & \num{113.7(1)} & \num{4.8(2)} & \num{1.6(2)} & \num{-6(1)} \\
		& $\rhombus$ & \num{115.3(4)} & \num{0.4(4)} & \num{-4(1)} & \num{3.5(4)} \\
		& $\pentago$ & \num{111.1(3)} & \num{1.1(8)} & \num{-3.5(6)} & \num{2.5(7)} \\
		\hline
		Figs. 3 and 4 & $\trianglepa$ / $\circlet$ & \num{110.9(5)} & \num{1(1)} & \num{-4.2(9)} & \num{3(1)} \\
		\hline
	\end{tabularx}
	\caption{Measured trap frequencies corresponding to the data sets shown in the main text. The geometric mean and the deviations from the mean trap frequency along the axes are given.}
	\label{tbl:trap_freqs}
\end{table*}

\subsection{Numerical simulations}
Throughout the presented work, numerical simulations of the physical system are performed using a numerical toolbox \cite{Antoine2014, Antoine2015}. We employ the included two-component framework to solve the coupled Gross-Pitaevskii (GP) equations including the two-component LHY contributions \cite{Petrov2015,Larsen1963}. In terms of the condensate wave functions $\Psi_i(\br) = N_i^{1/2}\psi_i(\br)$, the extended GP equations for components $i = 1, 2$ can be written as
\begin{align}
	\mu_i \Psi_i = \Big( - \frac{\hbar^2}{2m_i} \nabla^2 + V + \mu^{(i)}_{\rm{MF}} + \mu^{(i)}_{\rm{LHY}}\Big) \Psi_i.
	\label{eq:2-comp GPE}
\end{align}
Here, $\mu_i$ is the chemical potential, $V(\br) = m \omega_0^2 \br^2/2$ is the spherical  potential, and we include contributions from both mean-field interactions, $\mu^{(i)}_{\rm{MF}}$, and quantum fluctuations, $\mu^{(i)}_{\rm{LHY}}$.
For equal masses, $m = m_1 = m_2$, the mean-field contributions are given by
\begin{align}
	\mu^{(i)}_{\rm{MF}} = \frac{4 \pi \hbar^2}{m} \left( a_{ii} n_i + a_{ij} n_{j} \right),
\end{align}
and the contribution from quantum fluctuations can be written as
\begin{align}
	\begin{split}
		\mu_\text{LHY}^{(i)} = \frac{16\sqrt{2\pi}}{3} \frac{\hbar^2}{m} \sum_\pm \Bigg[ 
		\left( a_{ii} \pm \frac{a_{ii}^2n_i - a_{ii} a_{jj} n_j + 2 a_{ij}^2 n_j}{\sqrt{(a_{ii}n_i - a_{jj}n_j)^2 + 4a^2_{ij}n_i n_j}} \right) \\
		\times \left( a_{ii} n_i + a_{jj} n_j \pm \sqrt{(a_{ii}n_i - a_{jj}n_j)^2 + 4a^2_{ij}n_i n_j} \right)^{3/2} \Bigg].
	\end{split}
\end{align}

For the magnetic field dependence explored in Fig. 4 of the main text, $\delta a < 0$ for magnetic fields below 56.83(2) G, which results in $\mu_\text{LHY}^{(i)}$ acquiring a small imaginary component. As pointed out in Ref. \cite{Petrov2015}, the LHY energy is insensitive to small variations in $\delta a$, and in particular its sign. This has justified setting $\delta a = 0$ explicitly in previous work on self-bound droplets \cite{Petrov2015,Semeghini2018,Ancilotto2018}. In this work, we only consider the real part of  $\mu_\text{LHY}^{(i)}$ in the numerical simulations. Note that a consistent theory which avoids this issue was recently developed \cite{Hu2020}.

To include inelastic losses in the dynamical simulations we add imaginary loss terms of the form $-i\hbar(K/2)|\Psi|^4\Psi$ to the right hand side of the extended GP equations as previously employed in Refs. \cite{Semeghini2018,Ferioli2019,DErrico2019},
\begin{align}
	-\frac{i}{2}\left(K_{iii}|\Psi_i|^4 + \frac{2}{3} K_{iij} |\Psi_i|^2|\Psi_j|^2 + \frac{1}{3} K_{ijj} |\Psi_j|^4\right)\Psi_i.
\end{align}
Here, $K_{ijk}$ denotes the three-body loss coefficients and the factors in front of each term take into account how many atoms are lost by the given loss process. The three-body loss coefficients are given by
\begin{align}
	K_{iii} &= \frac{1}{3!} K^{\rm{th}}_{iii} \left[1+\frac{6}{n_i^2} \frac{\partial (E_{\rm{LHY}}/V)}{\partial g_{ii}}\right] \\
	K_{iij} &= \frac{1}{2!} K^{\rm{th}}_{iij} \left[1+\frac{2}{n_i^2} \frac{\partial (E_{\rm{LHY}}/V)}{\partial g_{ii}} + \frac{2}{n_i n_j} \frac{\partial (E_{\rm{LHY}}/V)}{\partial g_{ij}} \right],
	\label{eq:LHYloss}
\end{align}
where $K^{\rm{th}}_{ijk}$ denote the thermal three-body loss coefficients, $g_{ij} = 4\pi\hbar a_{ij}/m$,  and the terms involving the energy density $E_{\rm{LHY}}/V$ include beyond mean-field corrections to the three-body correlation functions of the mixture \cite{Cheiney2018}.

The loss coefficients involving atoms in the $\1$ state have previously been measured to be compatible with the $^{39}$K background value of \SI{7.74e-29}{cm^6/s} \cite{Cheiney2018, Zaccanti2009,Lepoutre2016} and we therefore use this value for $K^{\rm{th}}_{111}, K^{\rm{th}}_{112}$, and $K^{\rm{th}}_{122} $ throughout this work. For the loss channel involving three $\2$ atoms, we use $K^{\rm{th}}_{222} = \SI{5.4e-27}{cm^6/s}$ based on Ref. \cite{Semeghini2018} which is compatible with measurements in Ref. \cite{Cheiney2018}. Reference \cite{Semeghini2018} gives an uncertainty on the absolute value of a factor 2, and we therefore use this as an upper bound for the loss rate in Fig. 3 of the main text. For simulations excluding LHY terms in the GP equations, we also neglect the terms involving $E_{\rm{LHY}}/V$ in \Cref{eq:LHYloss}.

\subsection{Energy evolution throughout the experimental duration}
\begin{figure*}[]
	\centering
	\includegraphics[width=\linewidth]{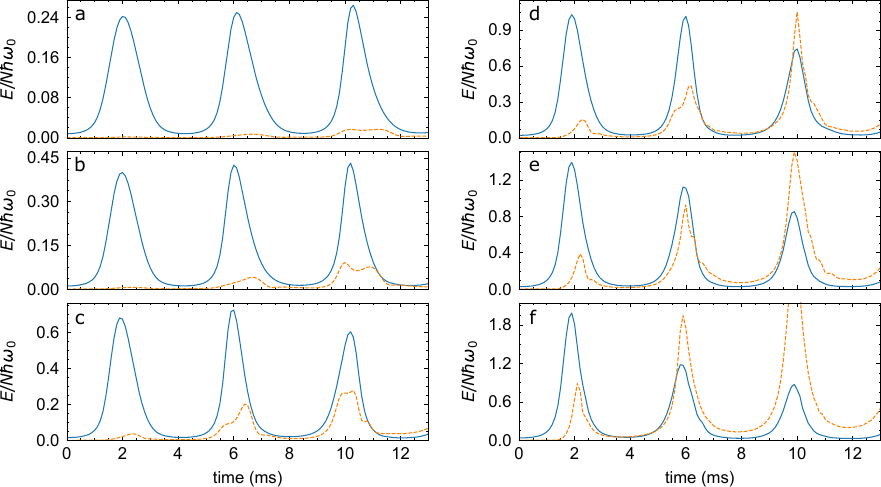}
	\caption{Simulated energy contributions from interactions throughout the experimental duration. The LHY and mean-field energies per particle are shown in blue and orange, respectively for $U = 0.15$ (a), $U = 0.27$ (b), $U = 0.58$ (c), $U = 1.2$ (d), $U = 2.2$ (e), and $U = 5.0$ (f).} 
	\label{fig:energy_evol}
\end{figure*}

Based on our simulation we can analyze the temporal evolution of the different energy contributions throughout the experimental duration. Figure \ref{fig:energy_evol} shows the LHY- and mean-field energies per particle for various values of $U$. Figure \ref{fig:energy_evol}(a) shows that the LHY-energy increases dramatically as a consequence of the increased density when the mixture contracts. The increasing density, however, also leads to enhanced inelastic losses resulting in a violation of the density ratio $n_2/n_1 = \sqrt{a_{11}/a_{22}}$ and in turn to an increasing energy contribution from mean-field interactions. Evidently, for low interaction strengths the system remains a LHY fluid throughout the entire experimental duration as shown for \Cref{fig:energy_evol}(a-c). For increasing $U$ the system deviates from a pure LHY fluid at earlier times due to the severe losses caused by the large amplitude oscillation. Nonetheless, the LHY-energy initially dominates the interactions for all interaction strengths and predominantly determines the oscillation frequency within the experimental duration.

\subsection{Data analysis}
We only include data sets where the oscillation can safely be characterized as monopolar based on separate fits to the oscillations along the two axes with Eq. (1) of the main text. We evaluate the frequency difference, $\delta \omega  = |\omega_y - \omega_z|$ and only include data sets where $\delta \omega = 0$ lies within a 2 $\sigma$ confidence limit. For data sets where this is fulfilled, we calculate the average of the measured radii $r = (r_y + r_z)/2$ for each evolution time and perform a fit with Eq. (1) of the main text which is then used to determine the experimental monopole oscillation frequencies.

The error bars in Fig. 3 of the main text are calculated as follows: Horizontal error bars are calculated by error propagation on $U$ and are dominated by the error on the total atom number. Since the atom numbers are measured independently before and after the data series, we have used the standard deviation on the atom number as a more conservative estimate than the standard error. Vertical error bars in Fig. 3 and Fig. 4 of the main text are calculated from the fit errors on the monopole oscillation frequencies and the error on $\omega_0$ which we determine by propagating the errors on $\omega_{i}'$.

\end{document}